\documentclass{article}
\usepackage[utf8]{inputenc}
\usepackage[style=numeric,sorting=none,citestyle=numeric-comp,backend=bibtex]{biblatex}
\usepackage{hyperref}
\usepackage{amsmath}
\usepackage{amssymb}
\usepackage{graphicx}
\usepackage{xcolor}
\usepackage{geometry}
\usepackage[toc,page]{appendix}
\title{Analytical computation of quantum corrections to non-topological soliton (bright soliton) within the saddle-point approximation}
\author{Andrei \;Kovtun$^{a,b}$\thanks{{\bf e-mail}:
akovtun@mpp.mpg.de}
\\
$^a${\small{\em
Max-Planck-Institut f{\"u}r Physik,}}\\
{\small{\em F{\"o}hringer Ring 6, D-80805, Munich, Germany
}}\\
$^b${\small{\em Arnold Sommerfeld Center, Ludwig-Maximilians-Universit{\"a}t,
}}\\
{\small{\em Theresienstra{\ss}e 37, 80333, M{\"u}nchen, Germany
}}\\
} 
\date{November 2020}

\begin{document}
\maketitle
\begin{abstract}
    Schr\"{o}dinger field theory with an attractive self-interaction possess non-topological extended solutions with a finite energy in both finite and infinite-volume cases, namely, bright solitons. The analytical form of the solution itself is well-known, though analytical investigation of the quantum fluctuations in this background still requires more thorough investigation, for instance, analytical computation of quantum corrections to this background within the saddle-point approximation. In the present work this gap is filled. Both 2-point Green's function and quantum corrections to the background are analytically computed and properly renormalized by means of momentum cut-off procedure. It is deduced that quantum corrections are indeed small provided that particle number is large. Also, we see that perturbation modes of continuum spectrum at bright soliton background generate a gap in the energy spectrum. Moreover, it turns out that the whole spectrum is continuous modulo zero-modes, which is similar to Sine-Gordon solitons.
\end{abstract}
\tableofcontents

\section{Introduction}
In quantum field theory two types of solitons are usually considered. The first type is topological solitons \cite{PhysRevD.10.4130,Polyakov:1974ek,Nielsen:1973cs,Abrikosov:1956sx}\footnote{And of course, all the other incalculable references listed for instance in \cite{manton_sutcliffe_2004}.}, which exist thanks to some non-trivial mapping of internal field space on coordinate space or space-time. For these solitons the quantization procedure is very well developed \cite{PhysRevD.11.1486,PhysRevD.11.2943,PhysRevD.12.1606} and both analytical and numerical studies of their quantum properties were carried out for different types of such solitons and within different models \cite{Dashen:1974cj,tHooft:1976snw,Ferreiros:2014mca,PhysRevD.66.125014,PhysRevD.68.045005}. The situation is more subtle with non-topological solitons, which can be stabilized by means of some conserved global current and correspond to fixed global charge \cite{Coleman:1985ki}. The procedure of quantizing non-topological solitons in the relativistic case was established in \cite{PhysRevD.13.2739}, but no actual computation according to this procedure was ever done\footnote{If we exclude Bose-Einstein condensates as non-topological finite energy solutions}, in spite of attempts given for instance in works like \cite{Rajaraman:1975qr}, where integration along symmetry direction was carried out. Nevertheless, it did not allow authors to go far beyond classical solution. Although, there are some works that address questions of fluctuating modes \cite{doi:10.1063/1.1665265,Gulamov:2013ema,Kovtun:2018jae}, neither analytical nor numerical computations of Green's function or quantum corrections\footnote{However, there is a work \cite{Graham:2001hr}, which claims computation of quantum corrections to the energy of Q-ball numerically. But for me and all my colleagues I have discussed this work with the results presented there are highly questionable. First of all, authors do not refer at all to original works by Friedberg, Lee, Sirlin, et al. and claim that they developed some other method, which anyway seems to be literally the same or provides minimal to none improvement compared to usual saddle-point technique. Secondly, one can see that eq. (11) in this work is simply incorrect, because it does not account for the mixing between complex field and its hermitian conjugate (look eq. \eqref{eq:time-independent Green's function} in my work). The fact that this system of equations describing perturbations in the Q-ball background can not be diagonalized by coordinate-independent transformation for inhomogeneous background is an easy fact to understand and this feature was considered in all the literature somehow related to perturbations at the top of non-topological configurations \cite{friedberg1976class,Kovtun:2018jae,Rajaraman:1975qr,Gulamov:2013ema,Smolyakov:2017axd,Panin:2018uoy}. Though maybe there was some trick done by the author, which was not mentioned at all. Thus, I must cast serious doubts about the results.} to the background of non-topological soliton like Q-ball by means of semi-classical methods was carried out. However, there are works \cite{Dvali:2015ywa} and \cite{Kanamoto_2003}, which suggests the procedure of non-perturbative diagonalization of non-relativistic Hamiltonian enclosed in finite volume, which implies cutting-off modes in the Hamiltonian leaving only those with momenta $p=\left\lbrace -2\pi/L,\,0,\,2\pi/L\right\rbrace$, thereby reducing infinite set of degrees of freedom to the finite one and then attempt to develop approximate quantization of the reduced Hamiltonian. As a result \cite{Dvali:2015ywa} yields interesting analytical results and \cite{Kanamoto_2003,Dvali:2013vxa} provide interesting numerical results. However, this method becomes increasingly demanding for a very large box or in case one wishes to include higher momentum modes making it harder to go to the infinite volume case. In addition, it is not sensible to ultraviolet modes because basically this method introduces explicit cut-off, which is not taken care of by renormalization, though this method is impressively useful in order to capture quantum effects qualitatively. 

The fact that equations for quantum fluctuations at the top of non-topological soliton are very difficult to solve does not come as a surprise, because non-topological solitons are time-dependent solutions and equations for their fluctuations are not diagonalizable by coordinate-independent transformation due to this feature. In this work I pursue a little bit more modest goal and compute all these things in the case of the non-relativistic field theory in 1+1 dimensions. In spite of the fact that it is hard to diagonalize equations of motion it is still possible to invert fluctuation operator, which is the inverse 2-point Green's function. Although, I must admit that this is manageable thanks to extreme simplicity of this system as it does not contain any square-intergable modes in the spectrum apart from non-oscillating zero modes, which basically means that no finite-energy modes are stuck inside potential well created by bright soliton, which is very similar to Sine-Gordon soliton \cite{Skyrme:1961vr}. In order to carry procedure of quantization out, I will make use of the identities found in \cite{PhysRevE.58.1064}, which involve actual eigenfunctions for continuum spectrum and zero-modes supplemented by additional ``quasi''-eigenfunctions, which make actual eigenfunctions up to completeness.

To make reading more coherent let me state here main results of this work:
\begin{enumerate}
    \item I compute 2-point Green's function of fluctuations in the background of bright soliton. This Green's function has poles at 
    \begin{equation}
        E_p =\omega +\frac{p^2}{2m},
    \end{equation}
    where $\omega$ is the frequency of bright soliton rotation in the internal space.
    \item We see that in order to excite a particle in the background of bright soliton within the sector with fixed charge we have to invest the energy equal to the gap $\omega$.
    \item Spectrum of quantum fluctuations in this background is substantially continuous modulo zero-modes.
    \item The classical energy of the bright soliton $E_{\text{cl.}}=-\frac{N}{384}\,m\,\alpha_{\mathrm{coll.}}^2$ is getting corrected
    \begin{equation}
       E= E_{\text{cl.}} + \delta E = -\frac{1}{384}\,m\,\alpha_{\mathrm{coll.}}^2\, \left(N-\left(5+\frac{16\pi^2}{15}\right)+\mathcal{O}(1/N)\right).
    \end{equation}
    where $\alpha_{\mathrm{coll.}} = N\,(\lambda/m^2)$ is the dimensionless collective coupling of internal degrees of freedom of bright soliton consisting of the particle number $N$, and dimensionless coupling of quartic self-interaction $\lambda/m^2$. Hence, it is apparent that correction scales as either $\mathcal{O}(\hbar)$ or as $\mathcal{O}(N^{0})$ compared to classical energy, and is small for large particle number $N\gg 1$.
\end{enumerate}
I will keep the proper explanation of these results to the main body and the conclusion.

The rest of the article is organized as follows: In Sec. \ref{sec:Classical theory} I present classical properties of bright solitons and connection to the relativistic theory, then in Sec. \ref{sec:quantum fluctuations} I introduce inverse Green's function of quantum fluctuations in the background of bright soliton and invert this operator in Sec. \ref{sec:greens function}, where I discuss back-reactions on the background and their compliance with the demand of fixed particle number, then in Sec. \ref{sec:Quantum corrections} quantum correction to the classical energy is computed.

\section{Classical theory}\label{sec:Classical theory}
 In the present work I consider Schr\"{o}dinger's field theory in 1+1 dimension confined in infinite spatial volume with an attractive self-interaction given by Lagrange density
\begin{equation}\label{eq:Lagrangian}
    \mathcal{L} = i \psi^{*}\dot{\psi}-\frac{1}{2 m}\left|\frac{d\psi}{dx}\right|^2 +\frac{\lambda}{8 m^2}\left|\psi\right|^4.
\end{equation}
This theory has internal global $U(1)$ symmetry
\begin{equation}
    \psi \rightarrow e^{i\alpha}\psi ,
\end{equation}
which provides conserved charge, namely, particle number
\begin{equation}
    N=\int dx\,|\psi|^2,
\end{equation}
for solutions vanishing at infinity or having periodic boundary conditions.

Varying action of this theory with respect to $\psi^*$ and $\psi$ one can deduce classical equations of motion
\begin{equation}\label{eq:classical EOM}
 i \partial_t \psi + \frac{1}{2m}\partial_{x}^{2}\psi+\frac{\lambda}{4 m^2}\left|\psi\right|^{2}\psi=0,  \\
\end{equation}
These equations posses well-known bright soliton solution
\begin{equation}\label{eq:classical solution}
    \psi_{\mathrm{cl}}(t,x) = e^{i\omega t} \sqrt{\frac{8 m^2 \omega}{\lambda}}\, \mathrm{sech}\left(\sqrt{2 m \omega}\,x\right),
\end{equation}
which has classical energy
\begin{equation}\label{eq:classical energy}
    E_{\mathrm{cl}} = -\frac{8\sqrt{2}}{3}\frac{(m\,\omega)^{3/2}}{\lambda},
\end{equation}
and corresponds to amount of particles
\begin{equation}\label{eq:particle number}
    N = \frac{8\sqrt{2 \,m \,\omega}\,m}{\lambda}.
\end{equation}
Therefore, it is indeed a non-topological soliton as it is a finite energy solution corresponding to fixed charge resulting from global current conservation.

Note that the energy of this bright soliton is negative and turns out to be binding energy as soon as we are working in the non-relativistic regime. However, we can make it up to the full energy adding the rest energy carried by $N$ quanta. Hence, the full relativistic energy is
\begin{equation}\label{eq:relativistic energy}
    E_{\mathrm{cl}}^{\mathrm{(rel.)}} = m\, N + E_{\mathrm{cl}} = \frac{8\sqrt{2}}{3}\frac{m}{\Tilde{\lambda}}\left( \left(\frac{\omega}{m}\right)^{1/2} - \left(\frac{\omega}{m}\right)^{3/2}  \right),
\end{equation}
where $\Tilde{\lambda} =\lambda/m^2$ is the dimensionless coupling constant. So, we see that $E_{\mathrm{cl}}$ is indeed just correction to the rest mass, because for non-relativistic approximation to hold inequality
\begin{equation}\label{eq:NR condition}
    m|\psi| \gg \left|\frac{\partial \psi}{\partial t}\right|
\end{equation}
must be satisfied, which results into
\begin{equation}\label{eq: omega smaller than m}
    \omega\ll m ,
\end{equation}
so we see that $\omega/m$ is another dimensionless small parameter. This is an important observation, because in non-relativistic theory there is no actual restriction for $\omega$ apart from its positiveness. But if we keep in mind relativistic set up, then it is evident, that $\omega$ has finite range and moreover must be small for non-relativistic approximation to hold.

Here we are finished with classical properties of bright soliton and can turn to study of quantum fluctuations in its background.
\section{Quantum fluctuations}\label{sec:quantum fluctuations}
Equation for the inverse Green's function in the background of bright soliton is easy to derive by taking second variational derivative of the action at the solution of classical equations of motion
\begin{equation*}
  \int d^2z \left. \frac{\delta^2 S}{\delta\psi_a (x)\delta\psi^{*}_c (z)}\right|_{\psi=\psi_{\mathrm{cl}}}\, G_{cb}(z;y)=i\delta_{ab}\delta^{2}(x-y),
\end{equation*}
where $x^{\mu}=(x^0,x^1)$, $\psi_a = (\psi,\psi^{*})^{T}$, $d^2z=dz^0\,dz^1$. Explicitly, this is
\begin{equation}\label{eq:Greens function equation (full)}
    \left(\begin{array}{cc}
     \displaystyle  i\partial_t + \frac{1}{2m}\partial_{x}^2 +\frac{\lambda}{2 m^2}\left|\psi_{\mathrm{cl}}\right|^2  & \displaystyle \frac{\lambda}{4 m^2}\psi_{\mathrm{cl}}^2 \\
      \displaystyle \frac{\lambda}{4 m^2}\psi_{\mathrm{cl}}^{*2}    & \displaystyle  -i\partial_t + \frac{1}{2m}\partial_{x}^2 +\frac{\lambda}{2 m^2}\left|\psi_{\mathrm{cl}}\right|^2 
    \end{array}\right)_{ac}\, G_{cb}\left(t,x;\tau,y\right)= i\delta(x-y)\delta(t-\tau)\delta_{ab}
\end{equation}

Off-diagonal elements are time-dependent. This can be removed by following transformation
\begin{equation}\label{eq:elimination of time-dependence}
    G(t,x;\tau,y)=T(t)\mathcal{G}(t-\tau;x,y)T^{\dagger}(\tau),
\end{equation}
where 
\begin{equation}
  T(t) =  \left(
    \begin{array}{cc}
        e^{i\omega t} & 0 \\
        0 & e^{-i\omega t} 
    \end{array}
    \right)
\end{equation}
is a unitary matrix.

Hence, this operator becomes just
\begin{multline}\label{eq:time-independent Green's function}
    \left(\begin{array}{cc}
     \displaystyle  i\partial_t-\omega + \frac{1}{2m}\partial_{x}^2 +\frac{\lambda}{2 m^2}f^2  & \displaystyle \frac{\lambda}{4 m^2}f^2 \\
     \\
      \displaystyle \frac{\lambda}{4 m^2}f^{2}    & \displaystyle  -i\partial_t-\omega + \frac{1}{2m}\partial_{x}^2 +\frac{\lambda}{2 m^2}f^2 
    \end{array}\right)_{ac}\, G_{cb}\left(t,x;\tau,y\right)=
    \\
    i\delta(x-y)\delta(t-\tau)\delta_{ab},
\end{multline}
where $f=f(x)=\left|\psi_{\mathrm{cl}}(x)\right|$.

Notice that $\mathcal{G}(t-\tau;x,y)$ is time-translation invariant, therefore we can perform Fourier transformation
\begin{equation}\label{eq:time Fourier}
   \mathcal{G}(t-\tau;x,y) = \int \frac{d\gamma}{2\pi}\,e^{-i\gamma (t-\tau)}\, \mathcal{G}(\gamma;x,y).
\end{equation}
and $SO(2)$ rotation
\begin{equation}\label{eq:transform to real basis}
    U\,\mathcal{G}(\gamma;x,y)\,U^{\dagger}=g(\gamma;x,y),
\end{equation}
where
\begin{equation}\label{eq:U}
  U =  \frac{1}{\sqrt{2}}\left(
    \begin{array}{rc}
        1 & 1 \\
        -1 & 1 
    \end{array}
    \right).
\end{equation}
This results to the equation for Green's function
\begin{equation}\label{eq:real, time-independent eq. for GF}
     \left(\begin{array}{cc}
     \displaystyle  -\omega + \frac{1}{2m}\partial_{x}^2 +\frac{3\lambda}{4 m^2}\,f^2  & \displaystyle -\gamma \\ \\
      \displaystyle -\gamma    & \displaystyle  -\omega + \frac{1}{2m}\partial_{x}^2 +\frac{\lambda}{4 m^2}f^2 
    \end{array}\right)_{ac}\, g_{cb}(\gamma;x,y)= i\delta(x-y)\delta_{ab}.
\end{equation}

Let me do a remark here. Notice that matrix $U$ defined in \eqref{eq:transform to real basis} is not actually a transformation to real basis it is just some transformation of complex valued  fields to some different complex valued ones.

If we plug the profile of bright soliton explicitly, one could easily recognize Hamiltonians coming from supersymmetric quantum mechanics. Unfortunately, their spectrum is of no use here, because equations are intertwined due to coordinate dependence and can not be diagonalized by coordinate-independent transformation.  Later we will give one more argument of futility of the eigenfunctions of modified P\"{o}schl-Teller potentials in this case.

I finish this section by writing down explicit expression for fluctuation operator
\begin{multline}
     \left(\begin{array}{cc}
     \displaystyle  -\omega + \frac{1}{2m}\partial_{x}^2 + 6 \omega \,\mathrm{sech}^2(\sqrt{2 m \omega}x)  & \displaystyle -\gamma \\ \\
      \displaystyle -\gamma    & \displaystyle  -\omega + \frac{1}{2m}\partial_{x}^2 +2 \omega \,\mathrm{sech}^2(\sqrt{2 m \omega}x) 
    \end{array}\right)_{ac}\, \mathcal{G}_{cb}(\gamma;x,y)=
    \\
    i\delta(x-y)\delta_{ab},
\end{multline}
and introducing convenient notations
\begin{equation}
     \left(\begin{array}{cc}
     \displaystyle  -\hat{L}  & \displaystyle -\gamma \\ \\
      \displaystyle -\gamma    & \displaystyle  -\hat{R}
    \end{array}\right)_{ac}\, \mathcal{G}_{cb}(\gamma;x,y)= i\delta(x-y)\delta_{ab},
\end{equation}
where 
\begin{equation}\label{eq:SS hamiltonians}
   \begin{array}{l}
       \hat{R} =  \displaystyle -\frac{1}{2m}\partial_{x}^2 - 2 \omega \,\mathrm{sech}^2(\sqrt{2 m \omega}x) +\omega , \\ \\
       
        \hat{L} =  \displaystyle -\frac{1}{2m}\partial_{x}^2 - 6 \omega \,\mathrm{sech}^2(\sqrt{2 m \omega}x) +\omega .
   \end{array} 
\end{equation}

In the next section I will invert this operator.
\section{Analytical form of Green's function}\label{sec:greens function}
\subsection{Computation}
Here we recall results from \cite{PhysRevE.58.1064} and consider specific functions, which can be turned to each other under the action of operators \eqref{eq:SS hamiltonians}.

First of all there are two sets of $L^{2}(\mathbb{R})$ functions satisfying following equations
\begin{equation}\label{eq:L^2 functions}
    \left\lbrace\begin{array}{rcl}
       \hat{R}\, r_{1}(x) & = & 0 \\
       \hat{R}\, r_{2}(x) & = & 2\omega\, l_{2}(x)
    \end{array}\right. ,\,\,\,\,\,
    \left\lbrace\begin{array}{rcl}
       \hat{L}\, l_{1}(x) & = & -2\omega\, r_{1}(x) \\
       \hat{L}\, l_{2}(x) & = & 0
    \end{array}\right.    .
\end{equation}
Apparently, only vector made of zero-modes $l_2$ and $r_1$ can solve the tricky eigenvalue problem
\begin{equation}\label{eq:eigenvalue problem of L^2 functions}
    \left(\begin{array}{cc}
     \displaystyle  -\hat{L}  & \displaystyle -\gamma \\ 
      \displaystyle -\gamma    & \displaystyle  -\hat{R}
    \end{array}\right)\left(\begin{array}{c}
         l (x)  \\ 
         r (x)
    \end{array}\right) = 0,
\end{equation}
posed by operator inverse to the Green's function. There is no any other linear combination, which can be built out of these functions to solve this eigenvalue problem.

However, there are continuum modes, which solve this eigenvalue problem:
\begin{equation}\label{eq:momentum modes}
    \left(\begin{array}{cc}
     \displaystyle  -\hat{L}  & \displaystyle -\gamma(p) \\ 
      \displaystyle -\gamma(p)    & \displaystyle  -\hat{R}
    \end{array}\right)\left(\begin{array}{c}
         l_{p} (x)  \\ 
         r_{p} (x)
    \end{array}\right) = 0,
\end{equation}
and corresponding eigenvalues are 
\begin{equation}
    \gamma(p) = -\left(\omega+\frac{p^2}{2m}\right).
\end{equation}
In spite all these functions do not solve the whole spectrum of the operator, they form a full set of functions satisfying following completeness relation
\begin{equation}\label{eq:completeness}
    \int\limits_{-\infty}^{+\infty}\frac{dp}{2\pi}\, r_{p}(x) l_{p}^{*}(y) + \sum\limits_{j=1}^{2}r_{j}(x) l_{j}^{*}(y) = \delta(x-y),
\end{equation}
which we will use to construct Green's function. This relation tells us that we cannot find anything more. Now, we can actually conclude that if we took linear combinations of eigenvalues of operators $\hat{L}$ and $\hat{R}$ we could not solve this problem, because they cannot be treated independently. Completness relation tells us that space of the functions we are considering is actually the space of one degree of freedom as it should be in case of non-relativistic field theory, if it were possible to diagonalize fluctuation operator and reduced it to two independent one-dimensional eigenvalue problems, that it would imply two propagating modes, which cannot be the case. There is no  doubling of degrees of freedom. Notice also that there are no square-integrable oscillating modes lying in $L^2(\mathbb{R})$, both square-integrable modes making up for completness are actually just some combinations of zero-modes, which makes this problem simple. This tells us that only continuum spectrum contributes to corrections to the energy to the contrary,for instance, to the case of kink.

Just for the purpose of convenience I will rewrite \eqref{eq:L^2 functions} and \eqref{eq:momentum modes} as
\begin{equation}
    \begin{array}{r}
       \hat{L}\, l_{\alpha}(x) = \lambda_{\alpha}^{l} r_{\alpha}(x)  \\
         \hat{R}\, r_{\alpha}(x)  = \lambda_{\alpha}^{r} l_{\alpha}(x)
    \end{array}.
\end{equation}
All these functions are explicitly shown in Appendix \ref{sec:eigenfunctions}.

Problem \eqref{eq:Greens function equation (full)} consists out of four equations, they can be split in two pairs. Let me show how to solve equations for first pair of Green's function components $g_{11}(\gamma;x,y)$ and $g_{21}(\gamma;x,y)$.

Equations for these two guys are
\begin{equation}\label{eq:GF left column}
    \left\lbrace\begin{array}{l}
       -\hat{L}\,g_{11}(\gamma;x,y)-\gamma\,g_{21}(\gamma;x,y)  = i\,\delta(x-y)  \\
       -\hat{R}\,g_{21}(\gamma;x,y)-\gamma\,g_{11}(\gamma;x,y)  = 0 \end{array}\right. .
\end{equation}
In order to solve these equation I assume Green's function to be
\begin{equation}
   \left\lbrace \begin{array}{l}
       g_{11}(\gamma;x,y) = \sum\limits_{\alpha}\left( s_{11,\alpha}^{ll}\,l_{\alpha}(x) l_{\alpha}^{*}(y) + s_{11,\alpha}^{lr}\,l_{\alpha}(x) r_{\alpha}^{*}(y) \right)  \\
        g_{21}(\gamma;x,y) = \sum\limits_{\alpha}\left( s_{21,\alpha}^{rr}\,r_{\alpha}(x) r_{\alpha}^{*}(y) + s_{21,\alpha}^{rl}\,r_{\alpha}(x) l_{\alpha}^{*}(y) \right) 
   \end{array} 
   \right.
\end{equation}
Also we can rewrite delta function at the RHS of \eqref{eq:GF left column} using completeness relation \eqref{eq:completeness}. Then, plugging all this into the first equation in \eqref{eq:GF left column} we get
\begin{equation}
    \sum\limits_{\alpha}\left( \left(-s_{11,\alpha}^{ll}\, \lambda_{\alpha}^{l}-\gamma\,s_{21,\alpha}^{rl} -
    i\right)\, r_{\alpha}(x)l_{\alpha}^{*}(y) + \left(-s_{11,\alpha}^{lr}\, \lambda_{\alpha}^{l}-\gamma\,s_{21,\alpha}^{rr} \right)\, r_{\alpha}(x)r_{\alpha}^{*}(y) \right) =0
\end{equation}
and also similar thing for the second equation in \eqref{eq:GF left column}. Demanding everything to vanish identically we deduce 12 equations defining unknown coefficients 
\begin{equation}
    \left\lbrace\begin{array}{rcl}
         -s_{11,\alpha}^{ll}\, \lambda_{\alpha}^{l}-\gamma\,s_{21,\alpha}^{rl} - i &=& 0  \\
         s_{11,\alpha}^{lr}\, \lambda_{\alpha}^{l}+\gamma\,s_{21,\alpha}^{rr} &=& 0 \\
         s_{21,\alpha}^{rr}\, \lambda_{\alpha}^{r}+\gamma\,s_{11,\alpha}^{lr} &=& 0  \\
         s_{21,\alpha}^{rl}\, \lambda_{\alpha}^{r}+\gamma\,s_{11,\alpha}^{ll} &=& 0
    \end{array}\right.,\,\, \alpha=1,2,p.
\end{equation}
Solving this linear system we get
\begin{equation}\label{eq:g11&g21}
    \begin{array}{rcl}
    -i\,g_{11}(\gamma;x,y) & = &\displaystyle \frac{2\omega}{\gamma^2}\,l_{2}(x)l_{2}^{*}(y)+\int\limits_{-\infty}^{+\infty}\frac{dp}{2\pi}\,\frac{\omega+\frac{p^2}{2 m}}{\gamma^2-\left(\omega+\frac{p^2}{2 m}\right)^2} l_{p}(x) l_{p}^{*}(y), \\
    \\
    -i\,g_{21}(\gamma;x,y) & = &\displaystyle -\frac{1}{\gamma}\,r_{1}(x)l_{1}^{*}(y)-\frac{1}{\gamma}\,r_{2}(x)l_{2}^{*}(y)+\int\limits_{-\infty}^{+\infty} \frac{dp}{2\pi}\,\frac{-\gamma}{\gamma^2-\left(\omega+\frac{p^2}{2 m}\right)^2} r_{p}(x) l_{p}^{*}(y),
\end{array}
\end{equation}
and repeating same computation for two remaining components we deduce
\begin{equation}\label{eq:g22&g12}
    \begin{array}{rcl}
    -i\,g_{22}(\gamma;x,y) & = &\displaystyle -\frac{2\omega}{\gamma^2}\,r_{2}(x)r_{2}^{*}(y)+\int\limits_{-\infty}^{+\infty}\frac{dp}{2\pi}\,\frac{\omega+\frac{p^2}{2 m}}{\gamma^2-\left(\omega+\frac{p^2}{2 m}\right)^2} r_{p}(x) r_{p}^{*}(y), \\
    \\
    -i\,g_{12}(\gamma;x,y) & = &\displaystyle -\frac{1}{\gamma}\,l_{1}(x)r_{1}^{*}(y)-\frac{1}{\gamma}\,l_{2}(x)r_{2}^{*}(y)+\int\limits_{-\infty}^{+\infty}\frac{dp}{2\pi}\,\frac{-\gamma}{\gamma^2-\left(\omega+\frac{p^2}{2 m}\right)^2} l_{p}(x) r_{p}^{*}(y).
\end{array}
\end{equation}
In the end, in order to complete the computation, a contour of integration must be specified. I will do $\gamma$ integration by means of Feynman prescription, namely, $\gamma^2-E_{p}^2+i\,0^{+}$.

Summarizing all the results we get
\begin{equation}\label{eq:full green's function}
    G(t,\tau|x,y) = \int \frac{d\gamma}{2\pi}\,e^{-i\gamma (t-\tau)}\,T(t)\, U^{\dagger}\,g(\gamma |x,y)\,U\, T(\tau)^{\dagger}
\end{equation}
\subsection{Gap in the continuum spectrum}
As the result of computation I deduced that propagator has poles at
\begin{equation}
    E_{p}=\omega +\frac{p^2}{2 m}.
\end{equation}
That is contrary to the vacuum dispersion relation. Nevertheless, this gap can be easily explained by means of following argument.

Solution we are studying lies in the sector of fixed charge $N$ (particle number). Therefore, any perturbation must not violate the number of particles which is conserved. Thus, in order to excite particle in the bright soliton background we must invest energy equal to $\omega$, that will correspond to the fact that we moved system to another sector with bright soliton composed out of $N-1$ particles and one external particle. Recall that for bright solitons holds following integral property
\begin{equation}
    \frac{dE_{cl}}{d\omega} = - \omega\,\frac{dN}{d\omega},
\end{equation}
so we can see that if we take one particle out of soliton it's energy is changed by
\begin{equation}\label{eq:energy shift by one particle}
    \Delta E_{cl} = \omega,
\end{equation}
that is exactly amount of energy which we need to excite a particle from continuous spectrum at the top of bright soliton. In other words we see, that energy of bright soliton with $N-1$ quantas is the same as energy of bright soliton with $N$ quantas and one additional quanta from free spectrum
\begin{equation}\label{eq:soliton + particle}
     E_{cl}(N-1)~=~E_{cl}(N)+E_{p=0}~=~
    E_{cl}(N)+\omega .
\end{equation}
The important lesson here is to keep in mind, that in this approach if one considers some scattering process with $m$ particles in the background of bright soliton, these particles will propagate in the background of bright soliton corresponding to the number of particles $N-m$. Therefore, we must remember that equalities \eqref{eq:energy shift by one particle} and \eqref{eq:soliton + particle} hold up to $1/N$ corrections and when we add one particle we change $\omega$ as well
\begin{equation*}
    \delta\omega =\frac{2 \omega}{N}.
\end{equation*}
Due to this fact approximation holds if we do not excite too many particles, which will result to the relative shift of $\omega$ larger than $1/N$.

\section{Quantum corrections}\label{sec:Quantum corrections}

Let me turn eventually to the computation of quantum corrections. In order to do that formally, I introduce partition function first

\begin{equation}\label{eq:generating functional}
    \mathcal{Z} = \int \mathcal{D}\psi\,\mathcal{D}\psi^{*}\, \exp\left(\displaystyle i\,\int dt \int dx\,\mathcal{L}\right) = \mathrm{tr}\left(e^{-i\,H\,T}\right)
\end{equation}
where Lagrangian is given by \eqref{eq:Lagrangian}.

Next we will restrict this functional to the configurations having fixed charge by plugging in projector
\begin{equation}
    \hat{P}(N)=\int\limits_{0}^{2\pi}\frac{d\alpha}{2\pi}\exp \left( -i\alpha \left(\hat{N}-N-\delta N \right) \right),
\end{equation}
where $\delta N$ is the renormalization of particle number. We need it here, because if we restrict put a restriction on time-ordered products (particle number in this case), they must be tuned to be finite or, in other words, regularized as any other measurable quantity.

After the inclusion of projector partition function becomes
\begin{equation}\label{eq:restricted Z}
    \mathcal{Z}_{N} = \int \mathcal{D}\psi\,\mathcal{D}\psi^{*}\, \int\limits_{0}^{2\pi}\frac{d\alpha}{2\pi}\, \exp\left(\displaystyle i\,S_{\mathrm{eff}}[\psi,\psi^{\dagger}]+i\alpha\, \left(N+\delta N \right)\right),
\end{equation}
where we have included in the action part of the projection operator
\begin{equation}
    S_{\mathrm{eff}}[\psi,\psi^{\dagger}] = \int dx\,dt\,\left(\mathcal{L}-\omega\, \psi^{\dagger}\psi\right),~~\omega = \frac{\alpha}{T},~~T=\int\limits_{-T/2}^{T/2} dt.
\end{equation}
Now we can employ saddle-point approximation with respect to both variables $(\psi,\psi^{\dagger})$ and $\alpha$. We shift them at the classical solution
\begin{equation}
    \left\lbrace
    \begin{array}{rcl}
       \psi(t,x)  & \rightarrow & \psi(t,x) +f(x)  \\
        \alpha & \rightarrow & \omega\,T+\alpha
    \end{array}\right.,
\end{equation}
where $f(x)$ is the modulus of \eqref{eq:classical solution} and $\omega$ is the same frequency as in the profile.

Hence, partition function becomes

\begin{multline}\label{eq:fixed N partition function}
    \mathcal{Z}_{N} = \int\mathcal{D}\psi\,\mathcal{D}\psi^{*}\, \int\limits_{-\omega T}^{2\pi-\omega T}\frac{d\alpha}{2\pi} \exp\left(-i\alpha\left(\int dx\,\left(f(x)\left(\psi+\psi^{\dagger}\right)+\psi^{\dagger}\psi\right)-\delta N\right)\right) \\
    \exp\left(i\int dt dx\int d\tau dy \frac{i}{2}\left(\psi^{\dagger}(\tau,y),\psi(\tau,y)\right)\hat{\mathcal{D}}(\tau-t|x,y)\left(\begin{array}{c}
         \psi(t,x)  \\
         \psi^{\dagger}(t,x)
    \end{array}\right)\right. + \\
    i\left.\int dt\,\left(\mathcal{L}_{\mathrm{int}}+\omega\delta \, N -\int dx\,\delta m\,f^2(x)\right)\right).
\end{multline}
In this expression for partition function
\begin{equation*}
    \hat{\mathcal{D}}(\tau-t|x,y)=\hat{\mathcal{D}}(t,x)\delta(\tau-t)\delta(x-y),
\end{equation*}
where $\hat{\mathcal{D}}(t,x)$ is the differential operator of linear fluctuations which we have specified in \eqref{eq:time-independent Green's function}, thereby it constitutes bi-linear part of Lagrange density. Then, we derive interaction terms, which are included in interaction density
\begin{equation}\label{eq:interaction density}
    \mathcal{L}_{\mathrm{int}}=\frac{(\lambda+\delta\lambda)}{8 m^2}\left(\psi^{\dagger}\psi\right)^2+\frac{(\lambda+\delta\lambda)\,f(x)}{4 m^2}\left(\psi+\psi^{\dagger}\right)\psi^{\dagger}\psi-\delta m \left( f(x)\left(\psi+\psi^{\dagger}\right)+\psi^{\dagger}\psi\right).
\end{equation}
Notice that I have not included $\delta N$ and $\delta m\,f^2$ here. The reason for that is that they contribute only in the renormalization of the energy of bright soliton, and are totally irrelevant for computation of diagrams. Also, at the level of the first quantum correction we do not need counterterm $\delta\lambda$, so, this is the only expression where I have put it just for the sake of generality. We will not need renormaliztion of coupling to compute first quantum correction.

Let me comment also on counterterm $\delta N$ constituting renormalization of particle number. From the first glance, one can say that we do not have this term for vacuum, which is not quite true. We need this to make sure, that at the level of operator-averages charge is properly normalized, namely, $\langle\hat{T}\lbrace\hat{N}\rbrace\rangle$ is UV-divergent due to time ordering, thus we need $\delta N$ in order to cancel this. In other words, application of the path integral formulation implies that we use time-ordered products, which contain divergences we have to take into account to get finite physical quantities.

Another important point is the integral over $\alpha$. It is evident that in order for the whole partition function $\mathcal{Z}_{N}$ to be non-zero, the argument of $\alpha$-dependent exponent must be exactly zero. From path integral definition of partion function with fixed particle number \eqref{eq:fixed N partition function} we deduce non-linear restriction on the averages of quantum operators
\begin{equation}\label{eq:charge conservation}
    \left\langle\hat{T}\left\lbrace
    \int dx\,\left(f(x)\left(\hat{\psi}+\hat{\psi}^\dagger\right)+\hat{\psi}^{\dagger}\hat{\psi}\right)-\delta N
    \right\rbrace\right\rangle = 0.
\end{equation}
Here $\hat{T}$ implies time-ordering. In the following we will use this condition to fix counterterm responsible for charge renormalization $\delta N$. This conditions is very important as it ensures conservation of the total charge. Here we compute everything by perturbation theory. Therefore, \eqref{eq:charge conservation} in the interaction representation picture becomes
\begin{equation}\label{eq:charge conservation in interaction picture}
    \frac{1}{\mathcal{Z}_{N}}\left\langle\hat{T}\left\lbrace
    \int dx\,\left(f(x)\left(\hat{\psi}+\hat{\psi}^\dagger\right)+\hat{\psi}^{\dagger}\hat{\psi}\right)
   \exp\left(i\int dt\,dx\,\mathcal{L}_{\mathrm{int}}\right) \right\rbrace\right\rangle = \delta N,
\end{equation}
where all the operators are in the representation of interaction, but I will not specify this explicitly by putting additional subscript at the field operators.

One can see from \eqref{eq:interaction density} that there are terms only proportional to positive orders of coupling constant $\lambda/m^2$, hence, perturbation theory in this case is valid. But if one would like to quantize this field canonically it is possible to impose condition \eqref{eq:charge conservation} on operators of the fluctuations. Therefore, one can see that it is necessary to non-linearly modify operators of fluctuations in order to impose conservation of particle number \cite{smolyakov2021nonlinear}.

Finally, everything is prepared in order to compute first correction to the energy of bright soliton, which is 
\begin{equation}\label{eq:total energy}
    E(N)=
    \lim_{T\rightarrow+\infty}\frac{i}{T}\log\frac{\mathcal{Z}_{N}}{\mathcal{Z}_{\mathrm{vac}}}.
\end{equation}
We will account corrections up to $\lambda$ order. Hence, energy is
\begin{multline}\label{eq:corrected energy}
    E(N)=E_{\mathrm{cl}}+ \\
    \frac{i}{T}\mathrm{tr}\left(\log\mathcal{D}(f(x))(\tau-t|x,y)-\log\mathcal{D}_{\mathrm{vac}}(\tau-t|x-y)\right)-\omega\,\delta N - \int dx\,\delta m\,f^2(x)+\mathcal{O}(\lambda) = \\
    E_{\mathrm{cl}}+\int \frac{dp}{2\pi}\frac{1}{2}\left(\gamma_{\mathrm{b.s.}}(p)-\gamma_{\mathrm{vac}}(p)\right)-\omega\,\delta N-\delta m\int dx\,f^2(x) +\mathcal{O}(\lambda),
\end{multline}
where classical energy is $\hbar$ independent and first correction is of order $\hbar$.

Let us start from evaluating energy coming from functional determinant. We do it by means of the same method as it was done for kink in \cite{weinberg_2012}. Namely, we assume for a moment that our system is enclosed to a very large box of size $L$ and quantize momenta. In order to do that let us compute asymptotics of functions $l_{p}(x)$ and $r_{p}(x)$
\begin{equation}
    \begin{array}{rcl}
        \displaystyle l_{p}(x) & \xrightarrow{x\rightarrow\pm\infty} & \exp\left(i\,px\pm\frac{i}{2}\delta(p)\right) ,\\
         \displaystyle r_{p}(x) & \xrightarrow{x\rightarrow\pm\infty} & \exp\left(i\,px\pm\frac{i}{2}\delta(p)\right)  ,
    \end{array}
\end{equation}
where shift of the phase is
\begin{equation}
    \delta(p)= -2\arctan\left( \frac{p\sqrt{2\omega/m}}{\omega-p^2/(2m)} \right).
\end{equation}
We see that both functions approach exactly the same phase shift simultaneously, reflecting the fact that they stand for one scattering mode, not two separate modes. Then, we impose quantization condition for momenta
\begin{equation}
    \begin{array}{lcll}
        p_{\mathrm{b.s.}}\,L +\delta\left(p_{\mathrm{b.s.}}\right) & = & 2\pi n, & \text{for bright soliton,} \\
         p_{\mathrm{vac}}\,L & = & 2\pi n, & \text{for vacuum modes,}  .
    \end{array}
\end{equation}
Hence we get that
\begin{equation}
    p_{\mathrm{b.s.}} = p_{\mathrm{vac}}-\frac{\delta(p_{\mathrm{vac}})}{L}+\mathcal{O}(L^{-2}).
\end{equation}
We plug this in
\begin{multline}
   \lim_{L\rightarrow+\infty} \frac{1}{2}\sum\limits_{n}\left( \gamma_{\mathrm{b.s.}}(p_\mathrm{b.s.})-\gamma_{\mathrm{vac}}(p_\mathrm{vac}) \right) = \\
    \lim_{L\rightarrow+\infty} \frac{1}{2}\sum\limits_{n}\left( \left( \omega+ \frac{p_\mathrm{b.s.}^2}{2\,m}\right)-\frac{p_\mathrm{vac}^2}{2\,m} \right) = \\
     \lim_{L\rightarrow+\infty}\left(\frac{L}{2}\sum\limits_{n} \frac{1}{L}\omega -\frac{1}{2\, L}\sum\limits_{n}\frac{ p_{\mathrm{vac}}\,\delta(p_\mathrm{vac})}{m} +\mathcal{O}(L^{-2})\right) = \\
     \frac{1}{2}\int dx\int\frac{dp}{2\pi}\,\omega - \int\frac{dp}{2\pi}\,\frac{p\,\delta(p)}{2m}.
\end{multline}
Notice that I used here substitution
\begin{equation*}
    \frac{1}{L}\sum_{n}\rightarrow\int\frac{dp}{2\pi}
\end{equation*}
in order to take infinite volume limit.

The resulting expression is divergent, but all these divergences are naturally canceled by counter--terms introduced earlier. Here I evaluate them explicitly.

First counterterm is mass renormalization coming from original Lagrangian
\begin{equation}
    \delta m \int dx\,f^2(x) = \frac{\lambda}{4\,m^2}\int \frac{dp}{2\pi}\, \int dx f^2(x) = 2\sqrt{2\omega/m} \int\frac{dp}{2\pi}.
\end{equation}

Now we come to the most interesting part, namely, evaluation of $\delta N$. One can see that in the condition \eqref{eq:charge conservation in interaction picture} two operators having different orders with respect to $\lambda$ are getting mixed. We evaluate this by perturbation theory using $\mathcal{L}_{\mathrm{int}}$ defined in \eqref{eq:interaction density}. Then condition for charge renormalization in the specified approximation becomes
\begin{multline}
    \delta N =\left\langle\hat{T}\left\lbrace
    \int dx\,\left(f(x)\left(\hat{\psi}+\hat{\psi}^\dagger\right)+\hat{\psi}^{\dagger}\hat{\psi}\right) \exp\left(i\int d\tau\,dy\,\mathcal{L}_{\mathrm{int}}\right)
    \right\rbrace\right\rangle = \\
    \int \left\langle\hat{T}\left\lbrace\hat{\psi}^{\dagger}(t,x)\hat{\psi(t,x)}\right\rbrace\right\rangle + \\
    \int dx \int d\tau dy\, f(x)\,f(y)\,\left(-i\,\delta m \right) \,\left\langle\hat{T}\left\lbrace\left(\hat{\psi}(t,x)+\hat{\psi}^{\dagger}(t,x)\right)\,\left(\hat{\psi}(\tau,y)+\hat{\psi}^{\dagger}(\tau,y)\right)\right\rbrace\right\rangle + \\
    \frac{i\,\lambda}{4 m^2}\int dx\int d\tau\,dy \, f(x)\,f(y)\left\langle\hat{T}\left\lbrace\left(\hat{\psi}(t,x)+\hat{\psi}^{\dagger}(t,x)\right)\left(\hat{\psi}(\tau,y)+\hat{\psi}^{\dagger}(\tau,y)\right)\hat{\psi}^{\dagger}(\tau,y)\psi(\tau,y)\right\rbrace\right\rangle +\\\mathcal{O}(\lambda).
\end{multline}
Here all the operators are in the interaction representation picture.

Now let me recall that I have computed Green's function explicitly for rotated variables, namely,
\begin{equation}
    \left(
    \begin{array}{c}
         \hat{\phi}_1  \\
         \hat{\phi}_2 
    \end{array}
    \right) = U\, \left(
    \begin{array}{l}
         \hat{\psi} \\
         \hat{\psi}^\dagger 
    \end{array}
    \right).
\end{equation}
Hence, I will evaluate $\delta N $ using these change of operators
\begin{multline*}
    \delta N = \int dx \int\frac{d\gamma}{2\pi} \frac{1}{2}\left(g_{11}(\gamma|,x,y)+g_{22}(\gamma|,x,y)\right) + \\
    i\,\int dx\,dy\,f(x)\,f(y)\,\left(g_{22}(0|x,y)\,\int \frac{d\gamma}{2\pi}\left(\frac{\lambda}{4 m^2}\left( \, g_{11}(\gamma|y,y)+3 \, g_{22}(\gamma|y,y) -2\delta m \right)\right)\right. + \\
    \left. 2 g_{12}(0|x,y) \int \frac{d\gamma}{2\pi}\,g_{12}(\gamma|y,y) \vphantom{\int}\right) = \\
    \frac{1}{2}\int dx\int\frac{dp}{2\pi} +\frac{1}{3}-\frac{16\pi^2}{45}.
\end{multline*}

I must draw reader's attention to the fact that here I do not account for non-oscillating parts of Green's function. These are used them only for inverting differential operator, but as soon as they do not oscillate, they do not contribute to propagation.

Now we can sum up everything in formula for corrections and see that all the divergences are cancelled and we are left with finite expression
\begin{multline}
    E(N)=E_{\mathrm{cl}} +  \\
    \frac{1}{2}\int dx\int\frac{dp}{2\pi}\,\omega - \int\frac{dp}{2\pi}\,\frac{p\,\delta(p)}{2m} + 2\sqrt{\frac{2\omega}{m}}\int \frac{dp}{2\pi} -\omega\left(\frac{1}{2}\int dx\int\frac{dp}{2\pi} +\frac{1}{3}-\frac{16\pi^2}{45}\right) +\mathcal{O}(\lambda) =\\
    -\frac{8\sqrt{2}}{3}\frac{(m\,\omega)^{3/2}}{\lambda} +\omega\left(\frac{5}{3}+\frac{15 \pi^2}{45}\right)   +\mathcal{O}(\lambda) .
\end{multline}
This is the final result. To make it more illustrative let's rewrite it in terms of collective coupling
\begin{equation}
    \alpha_{\mathrm{coll.}}=\frac{\lambda}{m^2} N = 8\sqrt{\frac{2\omega}{m}}.
\end{equation}
Re-expressing this through collective coupling, we get
\begin{equation}\label{eq:classical +quantum energy}
    E(N) = -\frac{1}{384}\,m\,\alpha_{\mathrm{coll.}}^2\left(N-\left(5+\frac{16\pi^2}{15}\right)+\mathcal{O}(1/N)\right).
\end{equation}
Therefore, we see that for sufficiently large $N$, quantum corrections are indeed just corrections to the classical energy of bright soliton, or to be more specific, if we consider non-relativistic field theory as the limit of the relativistic one, that would be the correction to the interaction energy of bright soliton constituents.

\section{Conclusion and Outlook}

To sum up, I have performed analytical computation of quantum fluctuations in the background of bright soliton \eqref{eq:g11&g21}, \eqref{eq:g22&g12}, \eqref{eq:full green's function}, which is the extended solution having finite energy in Schr\"{o}dinger field theory in 1+1 dimensions with attractive self-interaction. Firstly, I have inverted operator of quantum fluctuations or in other words computed 2-point Green's function of quantum fluctuations in the background of bright soliton. This propagator appeared to have poles in the continuum spectrum $E_p=\omega+p^2/(2m)$. The gap $\omega$ results from the constraint imposed by fixed particle number in the system and shows that if one excites a particle, the energy equal to the gap $\omega$ must be invested in order to tear the particle apart from bright soliton. Excitation of a single particle changes frequency as well, but only by amount of energy regulated by $1/N$ corrections. Therefore, one must keep track that the amount of quanta involved in the scattering process should be small compared to $N$ and if one studies some scattering process of $m$ particles this scattering happens in the background of bright soliton with $N-m$ quanta in it. Then I computed quantum corrections to the energy of this solution \eqref{eq:classical +quantum energy}. One can see that quantum corrections are controlled by $1/N$-expansion and indeed small for $N\gg 1$. 

In the end I would like to mention one more interesting observation. One can see that there are no integrable oscillating modes, this follows from the fact that only continuum spectrum solves the eigenvalue problem \eqref{eq:eigenvalue problem of L^2 functions} and all the others modes, which make up to completeness are actually just combinations of zero-modes of bright soliton. This suggest analogy with Sine-Gordon soliton and poses a question, if there is similar factorization of S-matrix for this model as it happens for Sine-Gordon model, and if there are objects dual to bright solitons.
\section*{Acknowledgment}
I would like to thank Gia Dvali, Mikhail Smolyakov and Emin Nugaev for useful discussions and reading the manuscript. Another thanks to E. Nugaev for suggesting the parallel between perturbation spectra of bright soliton and Sine-Gordon soliton.
\begin{appendices}
\setcounter{equation}{0}
\section{``Eigenfunctions'' and completeness relation}\label{sec:eigenfunctions}
\renewcommand{\theequation}{\thesection.\arabic{equation}}
We have two linear differential operators inside the fluctuation determinant, which are
\begin{equation}\label{eq:L and R operators}
   \begin{array}{l}
       \hat{R} =  -\displaystyle \frac{1}{2m}\partial_{x}^2 - 2 \omega \,\mathrm{sech}^2(\sqrt{2 m \omega}x) +\omega  \\ \\
       
        \hat{L} =  -\displaystyle \frac{1}{2m}\partial_{x}^2 - 6 \omega \,\mathrm{sech}^2(\sqrt{2 m \omega}x) +\omega 
   \end{array}. 
\end{equation}
These operators enjoy a set of function, which we will list below. First part of the ``spectrum'' consist of $L^{2}(\mathbb{R})$ functions
\begin{equation}\label{eq:L^2 eigenfunctions}
    \left\lbrace\begin{array}{rl}
        l_{1}(x) = & (2 m \omega)^{\frac{1}{4}}\left(1-\sqrt{2 m \omega}\, x \tanh{(\sqrt{2 m \omega}\, x)}\right)\mathrm{sech} (\sqrt{2 m \omega}\,x) \\
        l_{2}(x) = &  (2 m \omega)^{\frac{1}{4}} \tanh{(\sqrt{2 m \omega}\, x)} \, \mathrm{sech} (\sqrt{2 m \omega}\,x) \\
        r_{1}(x) = & (2 m \omega)^{\frac{1}{4}}\,\mathrm{sech} (\sqrt{2 m \omega}\,x) \\
        r_{2}(x) = & (2 m \omega)^{\frac{1}{4}}\sqrt{2 m \omega}\, x\,\mathrm{sech} (\sqrt{2 m \omega}\,x)
    \end{array}\right.,
\end{equation}
which obey following normalization conditions
\begin{equation}\label{eq: normalization for L^2 functions}
    \int dx \, r_{i}(x)l_{j}^{*}(x) = \delta_{ij},~~i,j=1,2
\end{equation}
and fulfill eigenvalue problem
\begin{equation}\label{eq:EV problem for L^2 functions}
    \left(\begin{array}{cc}
     \displaystyle  \hat{L}  & \displaystyle \gamma_{i} \\ 
      \displaystyle -\gamma_{i}    & \displaystyle  \hat{R}
    \end{array}\right)\left(\begin{array}{c}
         l_{i} (x)  \\ 
         r_{i} (x)
    \end{array}\right) = 0,
\end{equation}
with eigenvectors and eigenvalues
\begin{equation}\label{eq:eigenvalues of integrable functions}
  \Vec{h}_1 = \left( \begin{array}{c}
        l_1 (x)  \\
        r_2(x) 
   \end{array}\right)~~\text{with}~\gamma = 2\omega ,~~~~ \text{and}~~~~
     \Vec{h}_2 = \left( \begin{array}{c}
        l_2 (x)  \\
        r_1(x) 
   \end{array}\right)~~\text{with}~\gamma =0
\end{equation}
Also there are two functions comprising continuous spectrum, namely,
\begin{equation}\label{eq:continuous eigenfunctions}
    \left\lbrace
    \begin{array}{ll}
        l_{p} (x) = & \displaystyle \frac{e^{i p x}}{\sqrt{2 \pi }\left( \omega + \frac{p^2}{2 m} \right)}\,\left( -\omega - \frac{p^2}{2 m}  - \sqrt{2} i \sqrt{\frac{\omega}{m}}\,p \tanh{(\sqrt{2 m \omega}\, x)} + 2 \omega \tanh{(\sqrt{2 m \omega}\, x)}^2 \right) \\
        \\
        r_{p} (x) = &  \displaystyle \frac{e^{i p x}}{\sqrt{2 \pi }\left( \omega + \frac{p^2}{2 m} \right)}\,\left( \omega - \frac{p^2}{2 m}  - \sqrt{2} i \sqrt{\frac{\omega}{m}}\,p \tanh{(\sqrt{2 m \omega}\, x)}  \right)
    \end{array}
    \right. ,
\end{equation}
which satisfy different eigenvalue problem
\begin{equation}\label{eq:EV problem for continuous functions}
    \left(\begin{array}{cc}
     \displaystyle  -\hat{L}  & \displaystyle \omega +\frac{p^2}{2 m} \\ 
      \displaystyle \omega +\frac{p^2}{2 m}    & \displaystyle - \hat{R}
    \end{array}\right)\left(\begin{array}{c}
         l_{p} (x)  \\ 
         r_{p} (x)
    \end{array}\right) = 0,
\end{equation}
and obey momentum delta normalization
\begin{equation}\label{eq:normalization for continuous functions}
      \int dx \, r_{p}(x)l_{k}^{*}(x) = 2\pi\delta (p-k) .
\end{equation}

One can check that functions \eqref{eq:L^2 eigenfunctions} and \eqref{eq:continuous eigenfunctions} make up completeness relation
\begin{equation}
    \int\limits_{-\infty}^{+\infty}dp\, r_{p}(x) l_{p}^{*}(y) + \sum\limits_{j=1}^{2}r_{j}(x) l_{j}^{*}(y) = \delta(x-y)
\end{equation}
\end{appendices}
\printbibliography
\end{document}